# A Fast and Accurate Nonlinear Spectral Method for Image Recognition and Registration

Luciano da Fontoura Costa*

*Instituto de Física de São Carlos, Universidade de São Paulo, Av. Trabalhador São Carlense 400,
Caixa Postal 369, CEP 13560-970, São Carlos, São Paulo, Brazil*

Erik Bollt†

*Mathematics and Computer Science Department,
Clarkson University, PO Box 5815, Potsdam, NY 13699-5815, USA*

(Dated: 19th March 2006)

This article addresses the problem of two- and higher dimensional pattern matching, i.e. the identification of instances of a template within a larger signal space, which is a form of registration. Unlike traditional correlation, we aim at obtaining more selective matchings by considering more strict comparisons of gray-level intensity. In order to achieve fast matching, a nonlinear thresholded version of the fast Fourier transform is applied to a gray-level decomposition of the original 2D image. The potential of the method is substantiated with respect to real data involving the selective identification of neuronal cell bodies in gray-level images.

Pattern correlation and image registration represent essential operations in statistical physics and image analysis. Given a template reference, one is asked to find its related instances in a usually larger signal space. Often called *template matching*, and sometimes *image registration*, such a task has usually been understood and implemented in terms of the convolution or correlation operators (e.g. [1, 2, 3]). Thus, given a reference pattern $t$, the identification of its instances along the space $s$ is obtained by looking for maximum peaks of $R(t, s) = t \circ s$, namely the correlation between images $t$ and $s$. While fast, previous FFT based methods of registration have shown significant accuracy problems. Being a bilinear operator, the correlation is proportionally affected by the intensity of gray scale images $t$ and $s$. As a consequence, the presence in $s$ of even a single point of high intensity can produce a false matching, substantially reducing the discriminatory power of the whole approach. Indeed, even when the patterns in $s$ have similar height, the correlation tends to produce several significant false alarms. Figure 1 illustrates the rather limited power of correlation for pattern recognition. The result of the correlation between the template in (a) and the signal in (b) is presented in (c). It is clear that almost any pattern within a broad range of intensities can produce false alarms. Such a poor discrimination is ultimately a consequence of the tolerance allowed by the correlation to perturbations of the original patterns. Though such a tolerance is essential to cope with the many distortions found in experimental data, it should be kept to manageable levels so as not to compromise the overall discriminability.

The current article presents a non-linear spectral approach to template matching which provides an excellent balance between discriminability and level of toler-ance to distortions. The method involves the decomposition of the template and signal into binary subsignals, which are then strictly compared. In order to enhance speed, the comparisons are performed by using the Fourier transform in order to obtain the correlation between the quantized signals. Although a similar scheme has been previously applied to 1D biological sequence matching (e.g. [4, 5]), its generalizations to images and higher dimensional patterns are discussed in the present work.

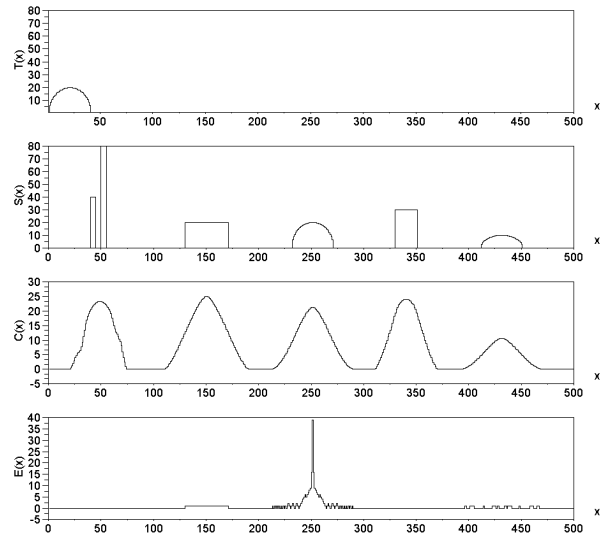

FIG. 1: The correlation between the template in (a) and the signal in (b), shown in (c), allows poor discrimination between the patterns. The spectral matching, shown in (d), provides excellent discrimination of the original template.

Initially, the image signals $t$ and $s$, originally with $G$ gray levels, have their intensity resampled to $g < G$ gray levels. This is illustrated in the diagram in Figure 2. Note that each of the resulting subsignals $t_i$ and $s_i$ are binary, in the sense of being composed exclusively by values zero or one. The strict comparison between each pair $(t_i, s_i)$, i.e. $C_i = t_i \odot s_i$ is performed by using the Fourier transforms. Let $T_i$ and $S_i$ be the respective Fourier transforms of $t_i$ and $s_i$. The strict matching between $t_i$ and $s_i$ can be conveniently obtained as the inverse Fourier transform of $TS$. Note that the peak value of $r_i$ will correspond to the maximum number of ones in the elementwise (Kronecker) product between $t_i$ and $s_i$ considering all possible lags [6]. The final result of the strict matching between $t$ and $s$ is obtained by adding all partial matchings, i.e. $M(t, s) = \sum_i C_i$. Therefore, the choice of the number of quantization levels $g$ provides the means for controlling the discriminative power of the method, with enhanced discrimination being obtained for larger numbers of quantization levels $g$. Figure 1(d) shows the strict matching of the patterns in (a) and (b) assuming 16 levels of binary quantization. The definite superiority of the proposed methodology can be readily verified by comparison with the traditional correlation n result shown in (c).

*Electronic address: luciano@if.sc.usp.br
†Electronic address: bolltem@clarkson.edu

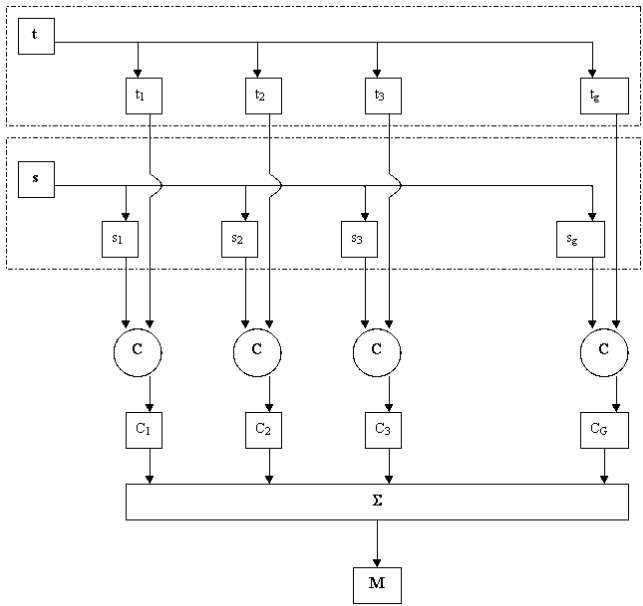

FIG. 2: Diagram presenting the main blocks in the strict pattern correlation method: the template and signal are quantized into $g$ binary subsignals which are subsequently compared by using the Fourier transform .

Next we turn our attention to the application of the strict matching methodology to patterns in the plane $(R^2)$. The methodology, which is a direct extension of the 1D matching strategy described above, therefore involves the following steps, given the specific number of quantizations levels $g$: (i) for $i = 1, 2, \ldots, g$ determine each thresholded version $s_i$ of the original image $S$ (with maximum gray level $G$) as $s_i = (1 + sign(S \geq i\Delta G))/2$, where $\Delta G = G/(g - 1)$; (ii) obtain the partial strict matchings, i.e. $C_i = t_i \odot s_i = F^{-1}(F(t_i)F(t_i))$, where $F(\cdot)$ is the Fourier transform; and (c) obtain the global strict match $M = \sum_{i=1}^{g} C_i$.

The potential of the proposed methodology is here illustrated with respect to real images of neuronal cell bodies (somata) obtained from the public *Brainmaps* dabatase [7]. Figure 3(a) shows a portion of the *Macaca mulatta* cortex (coronal slice), containing a number of Nissl stained neuronal cell bodies with approximate but varying sizes, shapes and gray level intensities. The neuronal cell body circled in Figure 3(a) has been chosen as the prototype (template) for the matching with the other cells. Note that this specific template cell presents some oblique elongation. The result obtained by applying traditional correlation followed by threshold (at half of the maximum gray level) and peak detection is shown in Figure éffig:exp(b). It is clear from this figure that the known shortcomings of the traditional correlation undermine the discrimination power of the method, resulting in the detection of almost every cell body in a way which is largely irrespective to their shape or gray-level. In addition, some cells have been missed while other structures have implied in overlapped peaks. The result of the strict matching considering the same template is shown in Figure 3(c). Note that only the cell bodies more similar to the template have in (a) in shape and gray level intensity have been mapped into well defined peaks in (b). Although remarkably good, this result can be further enhanced by thresholding the image in (b) at level $T$, detecting the obtained connected groups of points, and calculating their respective centers of mass. Figures 3(c) shows the obtained centers of mass superimposed onto the original image ($T$ at half of the maximum gray level value in (b) and $g = 4$). Only those cell bodies more similar to the template have been identified, without superposition. It is also clear from this result that, in addition to providing a good discrimination power, the strict matching also caters for reasonable level of tolerance to pattern alterations.

All in all, we have shown that high quality matching between patterns can be achieved by using a non-linear spectral method involving the decomposition of the signal into subsignals followed by their strict comparison. While such a methodology had been previously applied with encouraging success for the matching between biological sequences in bioinformatics, we have shown that its superiority is also manifest for the case of two- and higher dimensional patterns.

### Acknowledgments

Luciano da F. Costa thanks CNPq (308231/03-1) for sponsorship. Erik Bollt thanks the US National Science Foundation for its support under NSF grant DMS-

0404778.

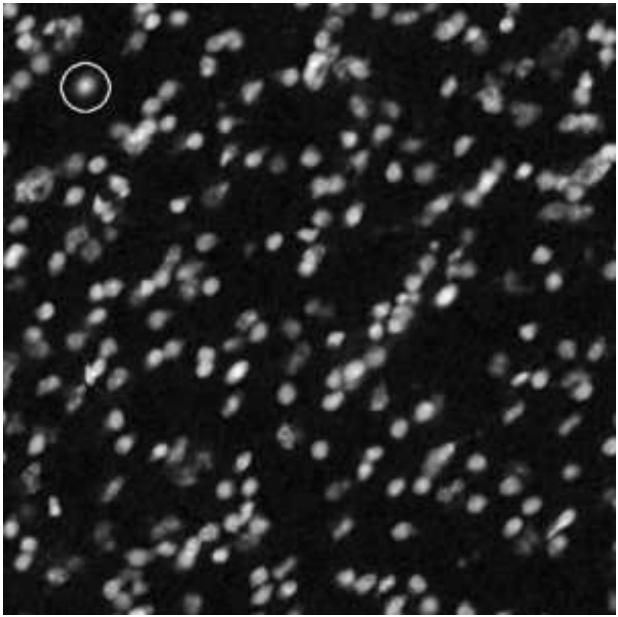

(a)

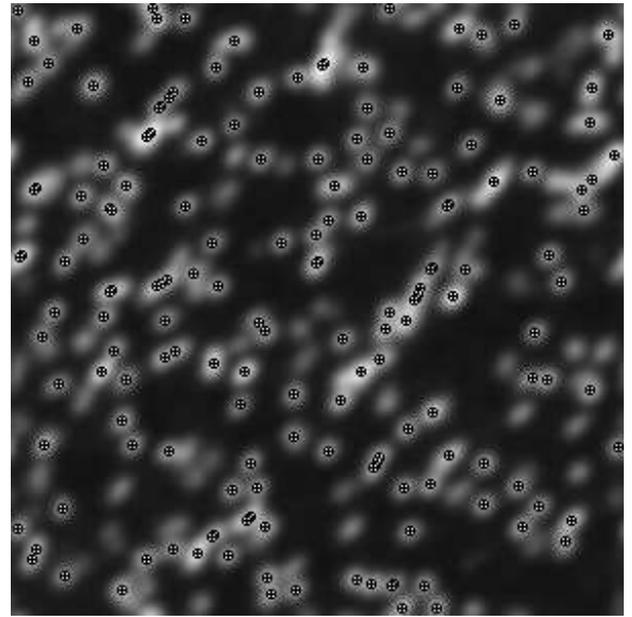

(b)

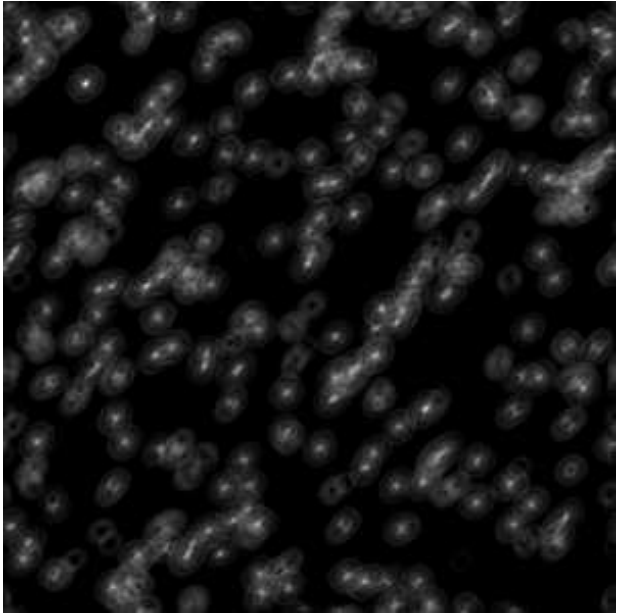

(c)

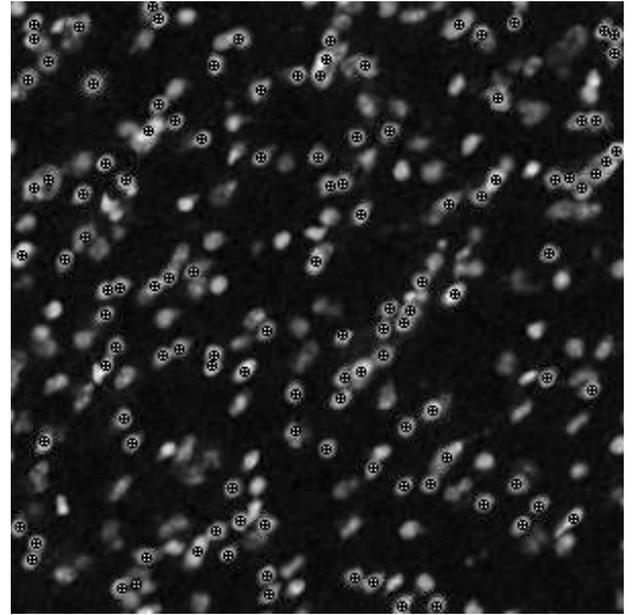

(d)

FIG. 3: Image of Nissl stained neuronal cells (*Macaca mullata*) (a). The template is shown inside the circle. The correlation between this template and the whole image is shown (b), and the strict matching results are shown in (c). The respectively detected cells superimposed onto the original image are presented in (d).